# Inverse spectroscopic optical coherence tomography (IS-OCT) for characterization of particle size and concentration

JAMES HOPE,[1,2] MATTHEW GOODWIN,[1,2] AND FRÉDÉRIQUE VANHOLSBEECK[1,2,*]

[1]*Dodd Walls Centre for Photonic and Quantum Technologies, Auckland 1010, New Zealand*
[2]*The Department of Physics, The University of Auckland, Auckland 1010, New Zealand*
*\*f.vanholsbeeck@auckland.ac.nz*

**Abstract:** Inverse spectroscopic optical coherence tomography (IS-OCT) methods apply inverse problem formulations to acquired spectra to estimate depth-resolved sample properties. In the current study, we modelled the time-frequency-distributions using Lambert-Beer's law and implemented IS-OCT using backscattering spectra calculated from Mie theory, then demonstrated the algorithm on polystyrene microspheres under idealized conditions. The results are significant because the method generates depth dependent estimates of both the concentration and diameter of scattering particles.

## 1. Introduction

In medicine and life sciences, the characterization of light scattering, absorption and polarization in biological tissues reveals information on their underlying structure and composition, allowing for diagnosis and monitoring in diverse applications such as arterial plaque in atherosclerosis, the size and nature of tumors, blood perfusion and oxygen saturation levels, and drug and nanoparticle delivery [1]. In particles, at the cellular and subcellular size-scale, light scattering and absorption are wavelength dependent (spectral) properties that can be described using Mie theory [2], a well-established stochastic model that forms the basis for the Monte Carlo model of multiple light-particle interactions [3-6]; alternatively, more recent scattering models also exist that include electromagnetic wave descriptions of light [7, 8].

In optical coherence tomography (OCT), low coherence spectroscopy (LCS) and low coherence interferometry (LCI), light from a low coherence source interact with a sample and a Michelson interferometer is used to extract depth-dependent information about the sample from the backscattered light [9]. In spectroscopic OCT (S-OCT) and LCS, broadband illumination and spectrally sensitive signal detection allows for characterisation of spectrally resolved absorption and scattering properties [6] which have been used to quantify media composition from absorption [10-14] and, when combined with Mie scattering models, to quantify particle size [4, 5, 14-21]. The use of inverse problem approaches to analyze particle size distribution in a sample is prevalent in systems using forward-scattering of light (see for example: [22, 23]) but is less common in backscattered light techniques such as OCT where it is known as inverse S-OCT (IS-OCT) [19, 20, 24-26].

In this paper, we briefly review the theory of S-OCT to show where each of the existing IS-OCT techniques are derived from [19, 20, 24-26] before extending the technique presented in [26], that used Lambert-Beer's law to describe the time frequency distributions, to backscattering spectra from Mie theory. Where these earlier techniques have demonstrated capability in estimating particle size [19, 20, 24-26], the technique presented here provides a quantitative estimate of particle size and concentration. Results are presented from data acquired on solutions of polystyrene microspheres with mean diameters of 0.17, 1.04 and 8.91 μm at concentrations of 0.156, 0.625 and 2.5 volume %. The results demonstrate the capability of IS-OCT to separate experimentally acquired spectra into several Mie scattering components.

*1.1 Theory*

1.1.1   S-OCT

In OCT, the fields originating from the sample arm $E_S$ and the reference arm $E_R$ interfere with one another to produce the interferogram detected by the photodetector, $I_D(k,z)$:

$$I_D(k,z) = \left\langle \left| E_S + E_S \right|^2 \right\rangle \quad (1)$$

where $k = 2\pi / \lambda$ is the wavenumber, $\lambda$ the wavelength, $2z$ the difference in optical path length between the two arms, and angled brackets indicate an average over the detector acquisition time which is assumed to be much longer than the time of an optical cycle. In a swept source OCT system, a laser with narrow linewidth, $\delta k$, spectrally sweeps across the light source bandwidth while the photodetector acquires the interferogram $I_D(k)$, which is then Fourier transformed, $\mathbb{F}$, to produce the sample depth dependent OCT signal $I_D(z)$ [6], Fig. 1a:

$$I_D(k) \overset{\mathbb{F}}{\Leftrightarrow} I_D(z) \quad (2)$$

Assuming that only light from single scattering events are present in the interferogram (first order Born approximation), that light attenuation is due to scattering and absorption (Lambert-Beer's law), and plane wave illumination from the light source, produces the general form for $I_D(z)$ of [28-30], Fig. 1b:

$$I_D(z) = \kappa \mu_b \exp(-2z\mu_t) h(z) \quad (3)$$

where $\kappa$ is a system constant which describes the influence of system components such as the photodetector, $h(z)$ is a function which describes the depth-dependent influence of system components such as beam divergence, $\mu_b$ the backscattering coefficient, and $\mu_t = \mu_s + \mu_a$ is the attenuation coefficient which is the sum of the scattering and absorption coefficients, $\mu_s$ and $\mu_a$.

Extraction and analysis of the depth-resolved spectral variation from the spectral interferogram is known as spectroscopic OCT (S-OCT), and is performed using a time-frequency-transformation (TFT) method such as a short-time Fourier transform (STFT), double window (DW), or the wavelet transform [6, 31]. In the case of STFT, a window, $w(k)$, is scanned across the interferogram and Fourier transformed to produce a set of intensity profiles, also called time-frequency distributions (TFD), which are specific to the spectral components encompassed by the window, $I_D[k,z;w(k)]$. By extracting depth indexed values from this set of TFDs, one constructs depth dependent spectra, $S_{TFD}[k,z;w(k)]$, Fig. 1c:

$$\mathrm{STFT}[I_D(k)] = \int_{-\infty}^{\infty} I_D(k') w(k'-k;\Delta k) \exp(-ik'z) \, dk' = S_{TFD}[k,z;w(k)] \quad (4)$$

where $w(k)$, is a window function with spectral width $\Delta k$. The process of windowing introduces a trade-off between the spectral and axial resolution in the resultant TFD due to the uncertainty principle, which states that sweeping a narrow window allows high spectral resolution but poor axial resolution and vice versa.

### 1.1.2 Parameters

Researchers have employed a range of experimental configurations using OCT to extract $\mu_t$, $\mu_a$, $\mu_s$ and $\mu_b$ from Eq. 3 for tissues and tissue phantoms [12, 13, 15, 29, 31], Fig. 1d. Once extracted, the spectral variation of $\mu_s$ and $\mu_b$ allows characterization of particle diameter using Mie theory [5, 16, 32], while the spectral variation of $\mu_a$ aids in tissue classification [31]. The parameters $\mu_a$, $\mu_s$ and $\mu_b$ are local material properties which, for identical scattering particles, can be calculated from the product of the number density of particles (with units: number of particles per unit volume), $\eta$, the cross-sectional area of the particle, $C$, and either the backscattering or scattering efficiency, $Q$ [2]:

$$\mu_x = \eta \sigma_x = \eta C Q_x \qquad (5)$$

where the subscript $x$ indicates either backscattering or scattering, and $\sigma$ is the effective cross section. $Q_x(\phi, \lambda, n_p, n_m)$ are functions of the particle diameter, $\phi$, light wavelength, $\lambda$, and the refractive indices of the particle and the medium, $n_p$ and $n_m$, and can be found using Mie theory [2, 33].

While it is not the focus of the current study, other methods have resolved the depth dependent spectra by TFT of the A-scan [31], $I_D(z)$, and then estimated the particle size from autocorrelation of the resultant spectra $S_{TFD}[k, z; w(k)]$ [4, 19, 34]. A further, alternative method for characterizing $\mu_s$ and $\mu_b$ is to describe spatial fluctuations in refractive index in the sample using the refractive index (RI) spatial auto-correlation function. In Refs. [24-26], the authors model this RI function using the Whittle-Matern family of functions that includes a shape factor determining the type of function, and then derive expressions for $\mu_s$ and $\mu_b$ using the shape factor among other parameters, Fig. 1e.

### 1.1.3 IS-OCT to estimate particle size distribution

The linear inverse problem $\mathbf{Ax} = \mathbf{d}$ can be used to estimate a sample's properties, $\mathbf{x}$, from experiment or modelling data, $\mathbf{d}$, via inversion of the coefficient matrix, $\mathbf{A}$, which relates the two. The Moore-Penrose pseudoinverse method for inversion of the coefficient matrix $\mathbf{A}$ is equivalent to the least squares method that minimizes the expression $\|\mathbf{Ax}_{MP} - \mathbf{d}\|_2$:

$$\mathbf{x}_{MP} = \mathbf{A}^\dagger \mathbf{d} = \mathbf{V}_P \mathbf{E}_P^{-1} \mathbf{U}_P^T \mathbf{d} \qquad (6)$$

where superscript T is the transpose, $\mathbf{E}$ is the singular value matrix, $\mathbf{V}$ and $\mathbf{U}$ are orthogonal matrices produced by singular value decomposition of $\mathbf{A}$, $p$ is the number of non-zero singular values and the subscript P indicates the first $p$ columns of the corresponding matrix

[37]. When negative values in **X** are not physically possible, a non-negative constraint can be imposed by changing negative values in **X** to 0. Tikhonov regularisation can stabilize the inverse problem by introducing a penalty term $\gamma \|\mathbf{x}_\gamma\|_2$ with regularisation parameter $\gamma$ and then minimizing $\|\mathbf{A}\mathbf{x}_\gamma - \mathbf{d}\|_2 - \gamma \|\mathbf{x}_\gamma\|_2$. Where in Tikhonov regularisation the L2 norm in the penalty term smooths the output by penalising large and sparsely distributed values in $\mathbf{x}_\gamma$, using the L1 norm on the penalty term promotes sparse solutions – i.e. those with several 0's in $\mathbf{x}_\gamma$ [38].

Researchers have applied this linear inverse problem formulation to IS-OCT using several different parameters and inversion algorithms: in Ref. [20], an iterative thresholding algorithm which minimizes the expression $\|\mathbf{A}\mathbf{x}_\gamma - \mathbf{d}\|_2 - \gamma \|\mathbf{x}_\gamma\|_1$, with the L1 norm implemented on the penalty term, was used to estimate diameters of microspheres from spectral interferogram data, Eq. 2, and a coefficient matrix that was populated with Mie spectra for different diameter particles each evaluated across different depths; in Ref. [21], k-means clustering was used to estimate diameters of microspheres from frequencies in the TFT of the depth interferogram $I_D(z)$ and a coefficient matrix that was populated using principal component analysis on a set of training data generated in controlled experiments; and, lastly, in Ref. [26], the Nelder-Mead algorithm, which is another iterative method, was used to estimate $\mu_b$ and the contributions of $\mu_a$ and $\mu_s$ to $\mu_t$ from the depth dependent spectra, $S_{\text{TFD}}[k, z; w(k)]$, Eq. 4, using the absorption spectra of chlorophyll obtained from literature and the Whittle-Matérn function derivations of scattering and backscattering.

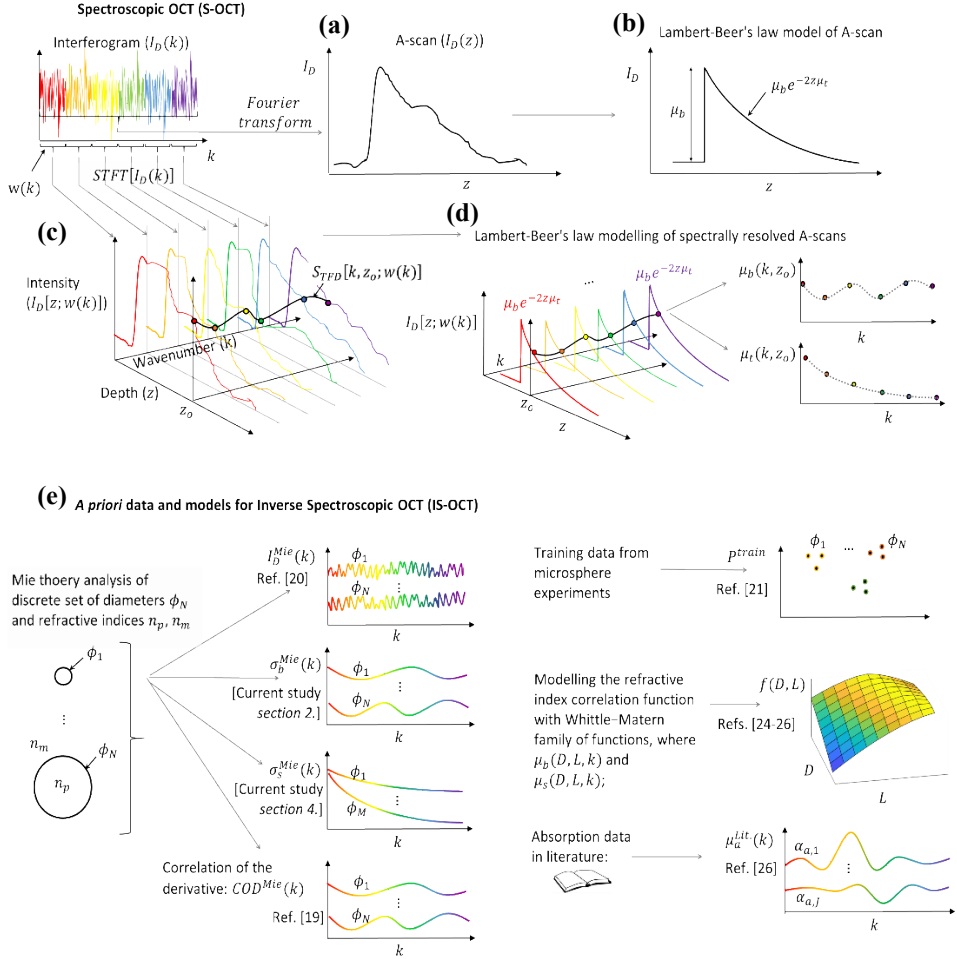

Fig. 1: (a) In OCT, an A-scan is produced by taking the Fourier transform of the entire bandwidth of the interferogram, producing high depth resolution but no spectral resolution. (b) Lambert-Beer's law can be used to extract backscattering and attenuation coefficients from an A-scan. (c) In S-OCT, a window slides across the interferogram and the STFT produces an A-scan for a narrow spectral band, shown here as different colored lines. The depth-resolved spectral profiles are then generated by extracting depth indexed values from this set of A-scans. (d) The spectrally resolved A-scans can also be modelled using Lambert-Beer's law, Eq. 4. (e) In IS-OOCT, discrete sets of a priori data from Mie scattering theory [19, 20], controlled experiments [21], and literature [26], as well as continuous (non-discrete) parameter functions [24-26] have been used to estimate sample properties through linear inverse problem algorithms.

## 2. Methods

### 2.1 Linear inverse problem formulation

For the current study, the linear inverse problem was formulated to estimate particle size and concentration from depth dependent spectra, $S_{TFD}\left[k, z; w(k)\right]$, Eq. 4, using a coefficient matrix that was populated with Mie backscattering efficiencies. Using the depth dependent spectra, as in Ref. [26], permitted incoherent averaging of multiple data sets to reduce speckle [35] without loss of signal amplitude – which can occur in the spectral interferogram method [20] if phase instability is present in the spectra, as is common in some wavelength-swept lasers.

In the current study, a set of TFDs were generated, and then each TFD was modelled using Lambert-Beer's law, Eq. 3, Fig 1c. Depth dependent spectra were then constructed from depth indexed values within the set of modelled TFDs, such that:

$$S_{TFD}[k, z; w(k)] = \kappa \mu_b(k, z_0) \odot \exp(-2z_0 \mu_t) \odot h(k, z_0) \quad (7)$$

where $z_0$ is an arbitrary depth, each variable is a $1 \times K$ vector of values, where $K$ is the number of TFDs in the set, and the operator $\odot$ indicates Hadamard multiplication and is equivalent to converting the vector of $K$ values produced in each function into a diagonal matrix and then performing matrix multiplication. When there are multiple scattering particles in the imaging volume, the voxel intensity of the imaging volume is the coherent sum of the backscattered spectra from each of the particles in the imaging volume [32], albeit with speckle. Therefore, for an imaging volume containing a set of particles with $N$ different diameters, the spectral modification by sample backscattering can be expressed as the sum of the backscattering coefficients from the particles in each diameter range:

$$\sum_{n=1}^{N} \mu_{b,n}(k, z_0) = \sum_{n=1}^{N} \eta_n \sigma_{b,n}(k, z_0) \quad (8)$$

where $\eta_n$ is the number density of diameter range $n$, $\sigma_{b,n}$ is the effective backscattering cross section for a single particle of diameter range $n$. Eq. 8 was converted to matrix notation, yielding:

$$\mu_b = \mathbf{A}\mathbf{x} \quad (9)$$

where $\mathbf{A}$ is a $K \times N$ matrix of backscattering cross sections for a single particle analyzed for $N$ different diameters that was populated using Mie theory, and $\mathbf{x}$ is a $N \times 1$ vector of number densities for each particle diameter. Eq. 9 was substituted into Eq. 7 and rearranged to solve for $\mathbf{x}$, yielding:

$$\mathbf{x} = \mathbf{A}^{\dagger} \left\{ S_{TFD} \odot h^{-1} \odot \left[ \kappa \exp(-2z_0 \mu_t) \right]^{-1} \right\} \quad (10)$$

where superscript † indicates the pseudoinverse.

## 2.2 OCT system

The OCT system used is polarisation sensitive with separate reference arms for horizontal and vertical polarisation and a quarter waveplate in the sample arm to circularly polarize the incident beam, and has been described in detail in an earlier study [39]. In the current study, this polarisation sensitivity was not utilised because the samples were not birefringent; the total back-reflected intensity was calculated as the sum of horizontal and vertical polarisation intensities. The OCT system used a swept source laser with 50 kHz sweep rate, that sweeps through approximately 1100 wavelengths spaced linearly in k-space (wavenumber space) across the 1259 to 1370 nm bandwidth (Axsun Technologies, AXP50125-6). At each wavelength, the interferograms for each polarisation were measured using balanced photodetectors (Thorlabs, PDB425C) and then digitized at 14 bits per sample and 125 MS/s (National Instruments, NI 5761). A galvo-mirror on the sample arm scanned the incident beam laterally across each sample using a 0.8 mm p-p amplitude, 6.25 Hz sawtooth wave while 8,000 A-scans were acquired at the laser sweep rate. This scanning approach produced uncorrelated speckle across the A-scans that was used to reduce speckle during data processing by averaging the resultant TFDs from a total of 80,000 A-scans [35]. The axial resolution was 10 μm in air and signal to noise ratio was 104 dB.

*2.3 Microsphere samples*

Microsphere samples were prepared from polystyrene particles with mean diameters of 0.17, 1.04 and 8.91 μm in a solution of diluted water and 0.02 % Sodium Azide (Spherotec, Inc., SPHERO™ Polystyrene Particles: PP-015-10, PP-10-10, PP-100-10), Fig. 2, diluted to the desired concentration with deionised water. After vortex stirring and then sonicating for 30 minutes, samples were prepared for each diameter at volume concentrations of 0.156, 0.625, and 2.5 %, Table 1. The concentrations were selected to be representative of organelles in biological cells, which for example can vary from 0.01 to 20 % in yeast [40], so that the results could aid future development of the method towards biological applications.

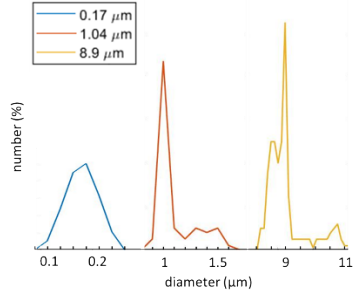

Fig. 2: The diameter distribution of polystyrene microsphere samples.

**Table 1. Properties of the microsphere samples used in the current study.**

| Concentration (volume %) | Mean microsphere diameter (μm) | | |
|---|---|---|---|
| 0.156 | 0.17 | 1.04 | 2.5 |
| 0.625 | 0.17 | 1.04 | 2.5 |
| 2.5 | 0.17 | 1.04 | 2.5 |

*2.4 Data processing*

TFDs were generated for each of the 80,000 interferograms by stepping a 20 nm Gaussian window at 0.1 nm increments across each interferogram and, at each step, zero-padding to 4096 points then applying the Fast Fourier Transform. The TFDs for each central wavelength were averaged to produce a single set of TFDs with reduced speckle. The power variation in the laser source spectra was characterised on a mirror covered with a diffusive filter. The TFDs from the mirror, $S_{TFD}[k, z_{mir}; w(k)]$, where $z_{mir}$ is the depth of the mirror, were then normalised and used to correct the microsphere data for source spectral variation, $H_{mir}(k)$, Fig. 3a. In addition, a calibration function, $H_{cal}(k)$, was defined using the desired spectral profile, $S_{MIE}[k, z_{mir}; w(k)]$, and the average spectra across the depth range of interest on the 0.17 μm 0.156 % microsphere sample, $S_{TFD}[k, z_{R1:R2}; w(k)]$, such that: $H_{cal}(k) = S_{MIE}[k, z_{mir}; w(k)] / S_{TFD}[k, z_{R1:R2}; w(k)]$. This calibration function was compared to the source spectral variation but not used to correct data because of the large presence of speckle. The calibration function was used in place of the system constant $\kappa$ because the latter could not be measured directly.

The confocal function, $h(k,z)$, was calculated for a Gaussian beam in water with waist radius of $r_w$ = 46 μm and focal depth $z_f$ = 1.263 mm, Fig. 3b, based on measurements of the OCT system. After multiplying data by $H_{mir}(k)$ and dividing by $h(k,z)$, the region of interest for analysis was visually selected to extend from the sample surface to the end of the linear region of signal decay on plots of several TFDs with logarithm intensity axis. The attenuation coefficient $\mu_t(k)$ was estimated as the gradient of a straight line fitted to the region of interest and was assumed to be independent of depth. The system constant, κ, was calculated as the product of the photodetector sensitivity (1.0 A/W) and the photodetector gain (2.5 × 10$^5$ V/A).

The attenuation coefficient $\mu_t(k)$ was estimated by fitting a straight line, using the MATLAB function *polyval,* to the logarithm of each of the TFDs, Fig. 3c; an approach made possible because the samples were considered homogeneous on the macro-scale. In a heterogeneous sample, the local attenuation coefficient can be estimated by dividing the intensity in the pixel of interest by the sum of the intensities of the pixels located at deeper depths [36].

Each column of the coefficient matrix **A** was populated with backscattering efficiencies that were generated using Mie theory equations implemented in MATLAB [33], the wavelength range of the source, of 1.259 to 1.370 μm (from product test sheet supplied by manufacturer), the refractive indices of water and polystyrene microspheres of 1.3225 and 1.59, respectively, and microsphere diameters in the range 0.1 to 10 μm. The wavelength range of each backscattering efficiency array was then clipped to 1.2771 to 1.3594 μm to account for the window size of the TFT and the data acquisition method that acquired the first 1024 points of the total 1120 points generated by the laser source.

In linear inverse problems, a small matrix condition number indicates good solution stability in the presence of noise in the experiment data. The backscattering efficiencies in the coefficient matrix spanned two orders of magnitude because of the large range of microsphere diameters, which produced a large condition number. Therefore, each column, containing the backscattering efficiencies for each diameter microsphere, were normalized against its maximum value, Fig. 3d, to reduce the matrix condition number from 8×10$^{15}$ to 8×10$^{12}$. Furthermore, the distance matrix, which calculates the Euclidean distance between each column, revealed that several of the backscattering efficiency arrays were similar to one another when observed across the narrow bandwidth of the laser source, particularly across 0.1 to 2 μm range that encompassed the 0.17 and 1.04 μm diameter samples, Fig. 3e. This minimal variation, which is observed as a low Euclidean distance values, acts to increase the matrix condition number and destabilize the solution. Therefore, the diameter range in the coefficient matrix was reduced to encompass the diameter ranges of the samples: 0.10, 0.14, 0.18, 0.22; 0.9, 1.0, 1.1, 1.2; and, 8, 8.4, 8.8, 9.2, 9.6, 10.0 μm. Columns with a Euclidean distance less than 1 were removed to reduce the matrix condition number to 2×10$^5$, which is comparable to the SNR of the TFDs, Fig. 3e. This targeted and restricted diameter range essentially limits the reconstruction to a categorization problem, where data are fit to different categories, though this may be improved with broader bandwidth sources.

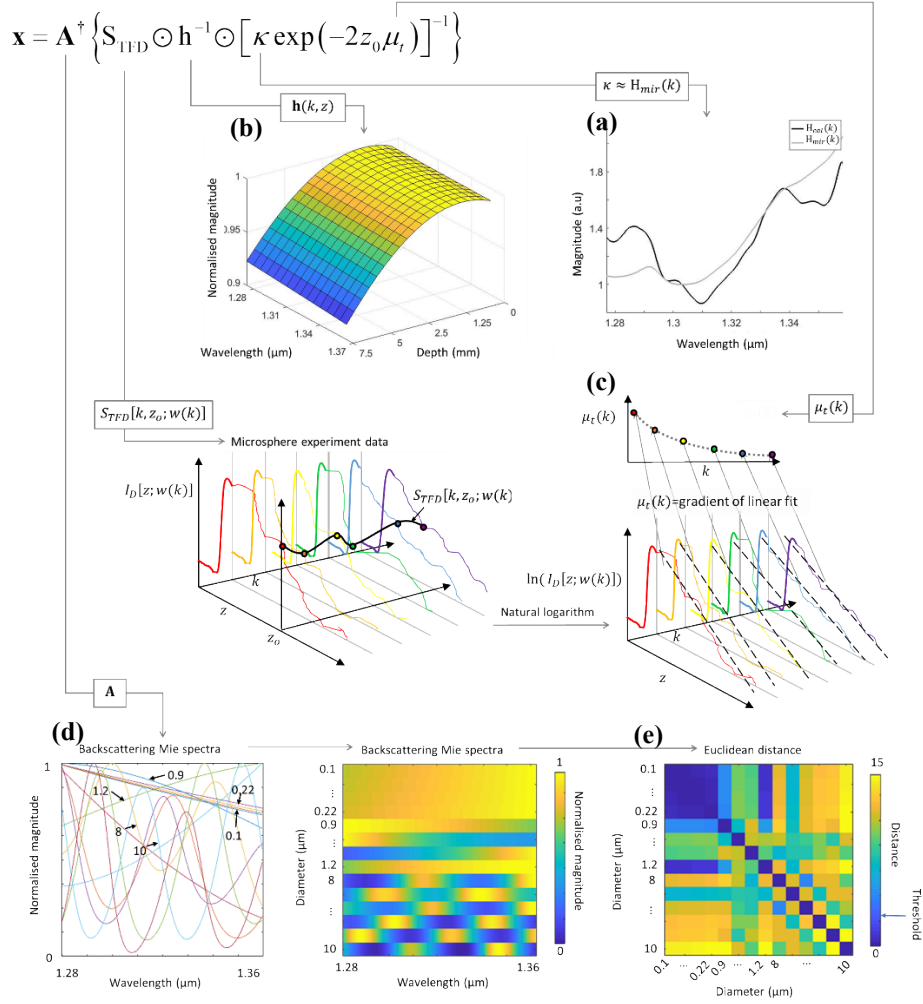

Fig. 3: Variables within Eq. 10. (a) TFDs from data collected on a mirror were normalized and used in place of the system constant $\kappa$. (b) The confocal function $h(k,z)$ was calculated for a Gaussian beam in water with waist radius of = 46 µm and focal depth = 1.263 mm. (c) The coefficient matrix was populated with backscattering efficiencies, calculated using Mie theory, for microsphere diameters: 0.10, 0.14, 0.18, 0.22; 0.9, 1.0, 1.1, 1.2; and, 8, 8.4, 8.8, 9.2, 9.6, 10.0 µm. (d) Backscattering efficiencies with a Euclidean distance less than 1 were removed to improve the condition number of the coefficient matrix to a value comparable to the SNR of the TFDs. (e) The attenuation coefficient $\mu_t(k)$ was estimated by fitting a straight line to the natural logarithm of each of the TFDs.

The depth dependent particle diameter distribution $\mathbf{x}(z)$ were estimated from the associated depth dependent spectra $S_{TFD}[k, z; w(k)]$ by solving Eq. 9 using the Lawson-Hanson algorithm [41]. The Lawson-Hanson algorithm iteratively finds a solution to the least squares expression with a non-negative constraint using an active-set method and is implemented in the MATLAB function *lsqnonneg*. In the current study, the least squares expression was $\min_{\mathbf{x}} \|\mathbf{Ax} - \boldsymbol{\mu_b}\|_2$, where $\boldsymbol{\mu_b}$ is the depth-dependent backscattering spectra extracted from experiment data and $\mathbf{Ax}$ is the backscattering spectra from Mie theory, Fig. 4.

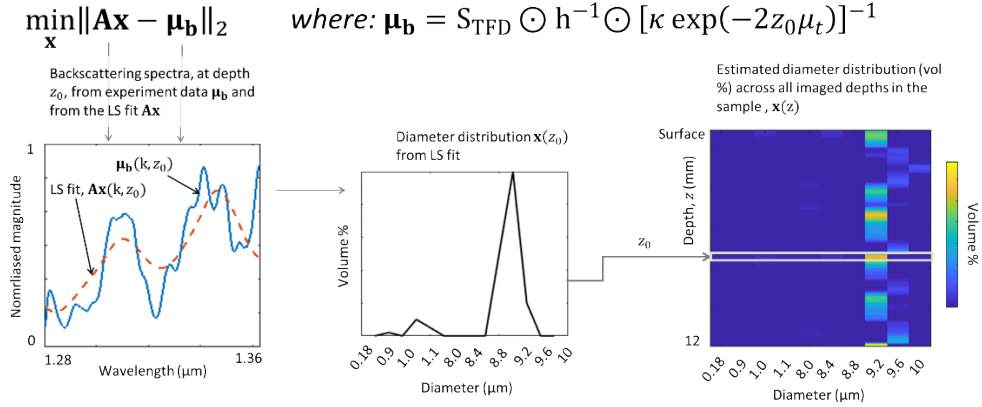

Fig. 4: The Lawson-Hanson algorithm iteratively changes the values in $\mathbf{x}$ to find a solution to the least squares expression $\min_{\mathbf{x}} \|\mathbf{Ax} - \boldsymbol{\mu_b}\|_2$ with a non-negative constraint, where $\boldsymbol{\mu_b}$ is the backscattering spectra calculated from the experiment data, and $\mathbf{Ax}$ is the backscattering spectra from Mie theory. The result is an estimate of the diameter distribution, $\mathbf{x}$, with the units of number density (number of particles per unit volume), which have been converted here to volume % for easier comprehension.

## 3. Results

The diameter distribution estimates from the surface to a depth of 12mm are presented in Fig. 5a for each of the nine microsphere sample preparations in Table 1. The diameter distribution is expressed as volume concentration (vol %) of each of microsphere diameters included in the coefficient matrix of backscattering Mie spectra: 0.18, 0.9, 1.0, 1.1, 8.0, 8.4, 8.8, 9.2, 9.6, and 10 µm. To aid quantitative analysis, the same data are also presented in box plots in Fig. 5b to provide the median, 25th and 75th percentiles of the diameter distribution estimates across depth range. In the box plots, diameter distributions are grouped into 3 diameter ranges of 0.18, 0.9:1.1, and 8:10 µm, that correspond to the diameter ranges of the microsphere samples.

In the three 0.17 µm sample preparations, of 0.156, 0.625 and 2.5 vol.%, the estimated diameter distributions contained a significant concentration of 0.18 µm microspheres across the 0 to 12 mm depth range but consistently estimated the concentrations to be lower than the actual sample preparation by a factor of approximately 4, Fig 5a (left column). A significant concentration of 1.1 µm microspheres and low concentration of large, 8 – 10 µm, microspheres was also common across the three 0.17 µm sample preparations, Fig. 5b (left column). This presence of larger microspheres in the diameter distribution is a possible cause of the underestimated concentrations of 0.18 µm microspheres. In the three 1.04 µm sample preparations, of 0.156, 0.625 and 2.5 vol.%, the estimated diameter distributions contained a significant concentration of 0.9, 1.0, and 1.1 µm microspheres across the 0 to 12 mm depth range, Fig. 5a (center column). The median concentrations in the 0.15 vol.%, 0.625 vol.% and

2.5 vol.% samples were, respectively, 0.03 vol.%, 0.3 vol.%, and 2 vol.%, Fig. 5b (center column), indicating relatively poor accuracy albeit an improvement on the 0.17 μm samples. In the three 8.49 μm sample preparations, of 0.156, 0.625 and 2.5 vol.%, the estimated diameter distributions contained significant concentrations of smaller microspheres and relatively low consistency across the 0 to 12mm depth range depths when compared to the 0.17 and 1.04 μm samples, Fig. 5a (right column).

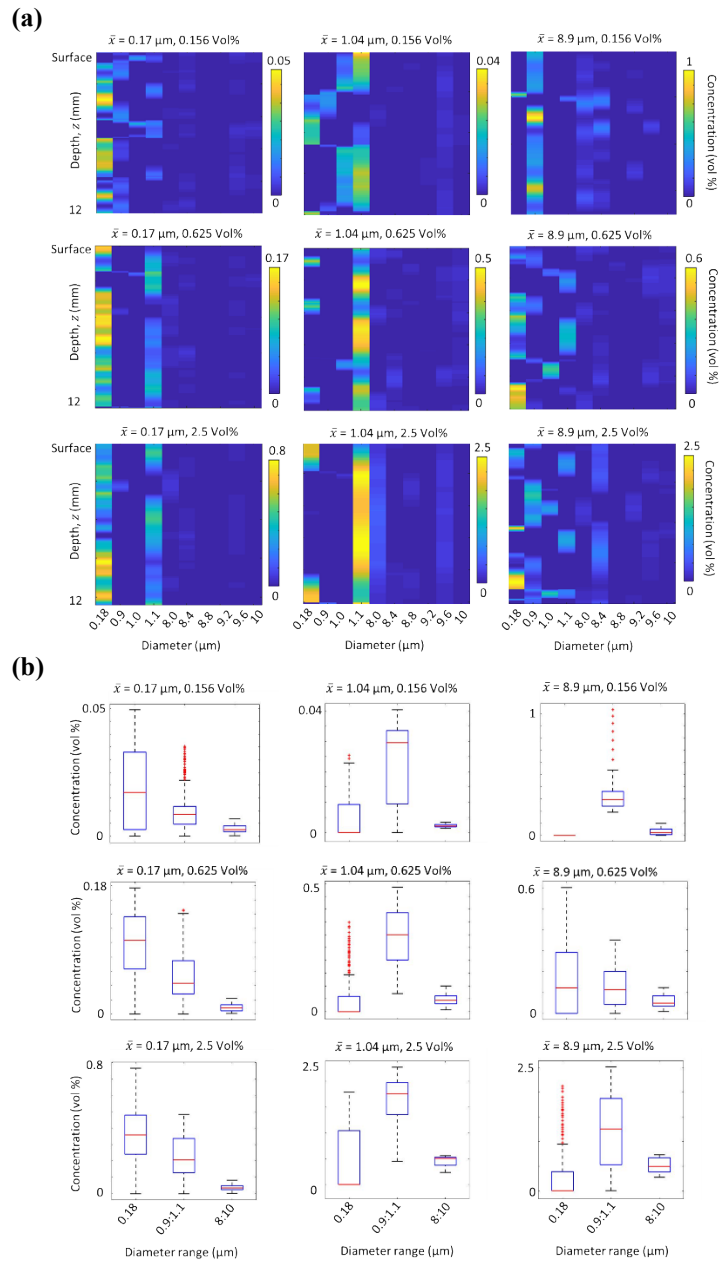

Fig. 5: Diameter distribution estimates for the nine microsphere sample preparations. (a) Color maps of the diameter distribution estimates with sample depth indicate good consistency in the

0.17 and 1.04 µm samples and poor consistency in the 8.49 µm samples across the 0 to 12 mm depth range. (b) Box plots showing the median, 25th and 75th percentiles of diameter distribution estimates across all depths show the concentrations were underestimated in all samples. In the box plots, data have been grouped into diameter ranges corresponding to the three microsphere samples.

## 4. Discussion

Mie scattering in OCT [15] and LCS [16] has previously been used to estimate particle size and concentration in samples without depth dependence, and in S-OCT to estimate the size of individual particles with depth dependence [5, 19, 20]. The current study has demonstrated the potential application of IS-OCT to quantify the depth dependent diameter and concentration of scattering particles using backscattering spectra from Mie theory. The results indicated reasonable accuracy in particle size estimation, good consistency of estimates with depth, and poor concentration accuracy for small particles (0.17 and 1.04 µm), but poor performance in these three areas for large particles (8.49 µm). In the nine sample preparations, the microsphere sizes and concentrations made it highly likely that multiple scatterers were present within each imaging volume (pixel). Incoherent averaging and data filtering approaches were essential to remove artefacts from system instability and speckle generated by multiple scatterers in the volume. Artefacts in the spectra are interpreted as scattering particles and are thought to be a significant contributor to errors in the results.

The limited variation in the Mie backscattering spectra for different particle diameters when observed across a narrow bandwidth hindered the inverse solution stability and was a major limitation of the method in the current study. Shorter wavelength and broader bandwidth light sources can reduce this instability by capturing more of the oscillations in the Mie backscattering spectra, Fig. 6, and may be achieved using visible wavelength and dual-band (visible and near infrared) OCT systems [42, 43].

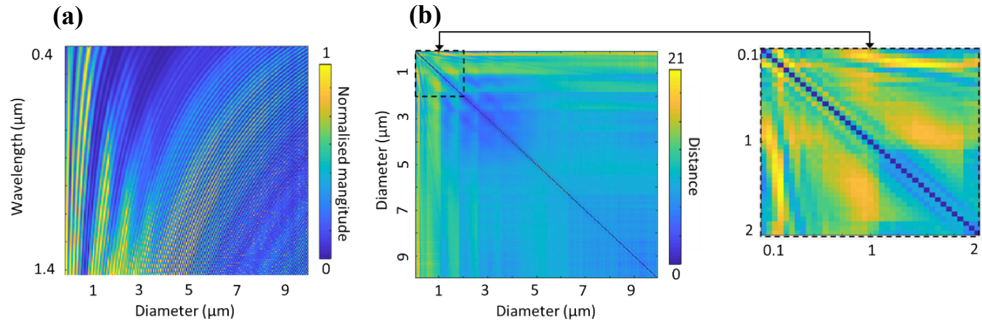

Fig. 6: (a) Backscattering spectra for polystyrene microspheres diameters 0.1 to 10 µm analyzed across a wavelength range of 0.4 to 1.4 µm and (b) the corresponding distance matrix showing that shorter wavelength and broader bandwidth light sources can improve solution stability by increasing the Euclidean distance and, in doing so, decreasing the coefficient matrix condition number.

With some prior knowledge of the sample, another method to improve solution stability is to remove some of the spectra, as was performed in the current study for the microsphere experiments. Prior knowledge of the sample can also be built into the algorithm in much the same way the $H_{mir}(k)$ was implemented; for example, a reference spectral profile might be generated from modelling or measurements across a population, and then spectral deviations from this reference used to indicate changes in size or concentration of scatterers in the sample, where high frequency deviations would indicate a change in large scatterers and low frequency changes a change in small scatterers. In Ref. [26], prior information was used to great effect to split the attenuation coefficient into absorption from chlorophyll and scattering from the sample structure. The frequency of oscillations in the Mie backscattering spectra can also be used to populate the coefficient matrix and is the basis for the interferogram IS-OCT implementation used in Ref. [20]. There, the authors accurately estimated the diameters of 1.0 and 4.5 μm polystyrene particles in air attached to glass slides though with deteriorating accuracy as the spheres occupied the same imaging volume – highlighting the problem of speckle in quantitative measurements as observed in the current study. The IS-OCT studies that fitted Whittle-Matern functions to the TFDs [24-26] did not describe speckle as a problem; possibly because the Whittle-Matern functions characterise the backscattering coefficient using a power law of the form $k^{4-D}$, where $D$ is the shape factor and is greater than zero, which limits the number of oscillations in the fitted curve. While this limitation is not a problem for small diameter microspheres [24], and biological tissues in mammal [24, 25] and coral [26], it may not extend well to large diameter scatterers or broader bandwidth OCT systems.

The IS-OCT problem becomes non-linear when $\mu_t$ cannot be extracted from the data or if non-linear effects such as multiple scattering events are included. Multiple scattering increases with the particle size and concentration. In these scenarios, non-linear formulation and inversion techniques are required, for which there are several techniques that are commonly applied to forward scattering problems [23].

Limitations of the current study were the limited number of diameters used in the coefficient matrix because of the long central wavelength and the narrow bandwidth laser source; the use of homogeneous microsphere samples meant that the depth dependent capabilities of the method were not demonstrated; and, multiple scattering events were not considered when extracting the attenuation coefficient. To improve the method in future, models of multiple scattering [15] might be used to improve estimation of the attenuation coefficient, and non-linear inverse problem methods [23] used to find a solution which extracts the backscattering, scattering and attenuation coefficients from the TFDs.

The approach to particle estimation presented in the current study, Eqs. 8, 9, and 10 can also be applied to the attenuation coefficient to estimate the composition of the media in the imaging volume. This application is described briefly here to highlight a potential basis of future studies. Assuming the contribution to the attenuation of light is the coherent sum of individual attenuation events, the spectral modification by sample attenuation can be expressed as the sum of the individual attenuation coefficients, which may be separated into contributions from scattering by particles and absorption by media. While Mie theory can be used to generate particle diameter dependent spectral absorption profiles, the diameter dependence is minor and so this is omitted in the current formulation and instead the imaging volume as a set of volume concentrations of different media using the relationship: $\mu_a(k) = 2.303 \tau \alpha_a(k)$, where $\tau$ is molar concentration, and $\alpha_a(k)$ is the molar extinction coefficient. Analysing for a set of $L$ different scattering particle diameters and $J$ different media absorption profiles yields:

$$\sum_{m=1}^{M} \mu_{t,m}(k, z_0) = \sum_{l=1}^{L} \eta'_l \sigma_{s,l}(k, z_0) + 2.303 \sum_{j=1}^{J} \tau_j \alpha_{a,j}(k, z_0) \quad (11)$$

where $\eta'_l$ is the number density of diameter range $l$, $\sigma_{s,l}$ is the effective scattering cross section for a single particle of diameter range $l$, $\tau_j$ is the molar concentration of media $j$, and $\alpha_{a,j}$ is the molar extinction coefficient for media $j$. In matrix notation, Eq. 11 can be expressed as $\mathbf{\mu_t} = \mathbf{By} + \mathbf{Cz}$, where $\mathbf{B}$ is a $K \times L$ matrix of scattering cross sections for individual particles of $L$ different diameters, $\mathbf{y}$ is a $L \times 1$ vector of number densities for each particle diameter, $\mathbf{C}$ is a $K \times J$ matrix of absorptivity profiles for $J$ different media, and $\mathbf{z}$ is a $J \times 1$ vector of volume concentrations for each of the defined media. The matrix $\mathbf{C}$ may be populated from existing absorptivity data on different biological tissues and other materials, for which the literature is extensive, while $\mathbf{B}$ may be populated using Mie theory. Eq. 11 can be rearranged to solve for $\mathbf{y}$ and $\mathbf{z}$ in a similar fashion to Eq. 10 after concatenating the coefficient matrices, $\mathbf{B}$ and $\mathbf{C}$, and the unknown parameter vectors, $\mathbf{y}$ and $\mathbf{z}$, into new matrices that can be solved using the matrix inversion methods described in *Section 1.1.3*:

$$\boldsymbol{\mu}_t = \mathbf{By} + \mathbf{Cz} = [\mathbf{B}, \mathbf{C}][\mathbf{y}, \mathbf{z}]^T$$
$$\therefore [\mathbf{y}, \mathbf{z}]^T = [\mathbf{B}, \mathbf{C}]^\dagger \boldsymbol{\mu}_t \tag{12}$$

## 5. Conclusion

A S-OCT method was presented that estimates depth-dependent particle size and concentration from TFDs and Mie backscattering spectra and is most suited to occurrences where multiple scatterers in the imaging volume prohibit IS-OCT analysis of individual particles. The method was demonstrated on polystyrene microspheres in solution using a coefficient matrix containing a limited set of backscattering spectra to compensate for system errors and narrow source bandwidth.

**Funding.** Marsden Fund and Royal Society of New Zealand (UoA1509)

**Disclosures**. The authors have no relevant financial interests in this article and no potential conflicts of interest to disclose.

**Data availability**. Data underlying the results presented in this paper are not publicly available at this time but may be obtained from the authors upon reasonable request.